\documentclass[12pt]{iopart}
\usepackage{iopams}  
\usepackage{graphicx}
\usepackage[dvipdfm]{color}
\usepackage[dvipdfm,bookmarks=true,bookmarksnumbered=true,bookmarkstype=toc]{hyperref}
\renewcommand{\mr}[1]{\mathrm{#1}}

\begin{document}

%{{{ Title

\title[]{Length sensing and control strategies for the LCGT interferometer}

\author{Y. Aso$^1$, K. Somiya$^2$, O. Miyakawa$^3$ for the LCGT Collaboration}

\address{$^1$ Department of Physics, University of Tokyo, Hongo 7-3-1,
Bunkyo-ku, Tokyo, Japan}
\address{$^2$ Graduate School of Science and Technology, Tokyo Institute of Technology,
2-12-1 Oh-okayama, Meguro-ku, Tokyo, 152-8551, Japan}
\address{$^3$ Institute for Cosmic Ray Research, University of Tokyo,
5-1-5 Kashiwa-no-Ha, Kashiwa City, Chiba, Japan}
\ead{aso@granite.phys.s.u-tokyo.ac.jp}

%}}}

%{{{ Abstract

\begin{abstract}
The optical readout scheme for the length degrees of freedom of the LCGT
 interferometer is proposed. The control scheme is compatible both with
 the broadband and detuned operations of the
 interferometer. Interferometer simulations using a simulation software
 Optickle show that the sensing noise couplings caused by the feedback
 control can be reduced below the target sensitivity of LCGT with the
 use of feed forward. In order to improve the duty cycle of the
 detector, a robust lock acquisition scheme using auxiliary lasers will
 be used.

\end{abstract}

\submitto{\CQG}
%\maketitle

%}}}

%{{{ \section{Introduction}

\section{Introduction}
The second generation interferometric gravitational wave detectors, such
as LCGT, advanced LIGO and advanced Virgo are planned to start
observations around 2017 to 2018.  The ambitious sensitivity goals set
by those projects require that the mirrors of the interferometers be
controlled with extremely high accuracy and lowest possible
disturbance. All of the above mentioned detectors will use an optical
configuration called Resonant Sideband Extraction
(RSE)\,\cite{mizuno_resonant_1993}, where a mirror is placed at the
anti-symmetric port of the interferometer to modify the spectral shape
of the quantum noises. As a result, the number of degrees of freedom
(DOFs) to be controlled is increased from the first generation
detectors.  Moreover, the strong laser power circulating in the
interferometer creates optical springs, resulting in opto-mechanical
couplings between otherwise independently suspended mirrors. This
complicated situation calls for a highly sophisticated design of the
interferometer sensing and control system.

In this paper, we will explain the signal extraction scheme to be used
in LCGT, the next generation gravitational wave detector in
Japan\,\cite{kuroda_status_2010}.  The focus of this paper is on the
length sensing and control. However, the control of the mirror
orientations is equally important and it shall be explained elsewhere in
a separate paper.

%}}}

%{{{ \section{LCGT interferometer}

\section{LCGT interferometer}

% \begin{figure}[tbp]
% \begin{minipage}{9cm}
%     \includegraphics[width=9cm]{Figure1.eps} \caption{Interferometer
% configuration and the sideband resonant conditions of LCGT. The laser
% beam is injected through the mode cleaner (MC) from the left of the
% figure. ETM: End Test Mass, ITM: Input Test Mass, BS: Beam Splitter,
% PRM: Power Recycling Mirror, SRM: Signal Recycling Mirror, OMC: Output
% Mode Cleaner. PR2, PR3, SR2 and SR3 are names of the folding
% mirrors. REFL, POP and AS\_DC are names of the signal extraction ports.}
% \label{Fig:InterferometerConfiguration}
% \end{minipage}
% \begin{minipage}{7cm}
% \makeatletter
% \def\@captype{table}
% \makeatother
%    \begin{tabular}{|c|c|}
% \hline
% Arm Cavity Finesse&1550\\ \hline
% Power Recycling Gain&10\\ \hline
% SRM Reflectivity&85\%\\ \hline
% SRC Detuning for DRSE&3.5$^\circ$\\ \hline
% Input Laser Power&51\,W\\ \hline
% Main Mirror Mass&22.8\,kg\\ \hline
%   \end{tabular}
% \caption{Interferometer parameters of LCGT.}
% \label{IFO Parameters}
% \end{minipage}
% \end{figure}

\begin{figure}[tbp]
\begin{center}
 \includegraphics[width=12cm]{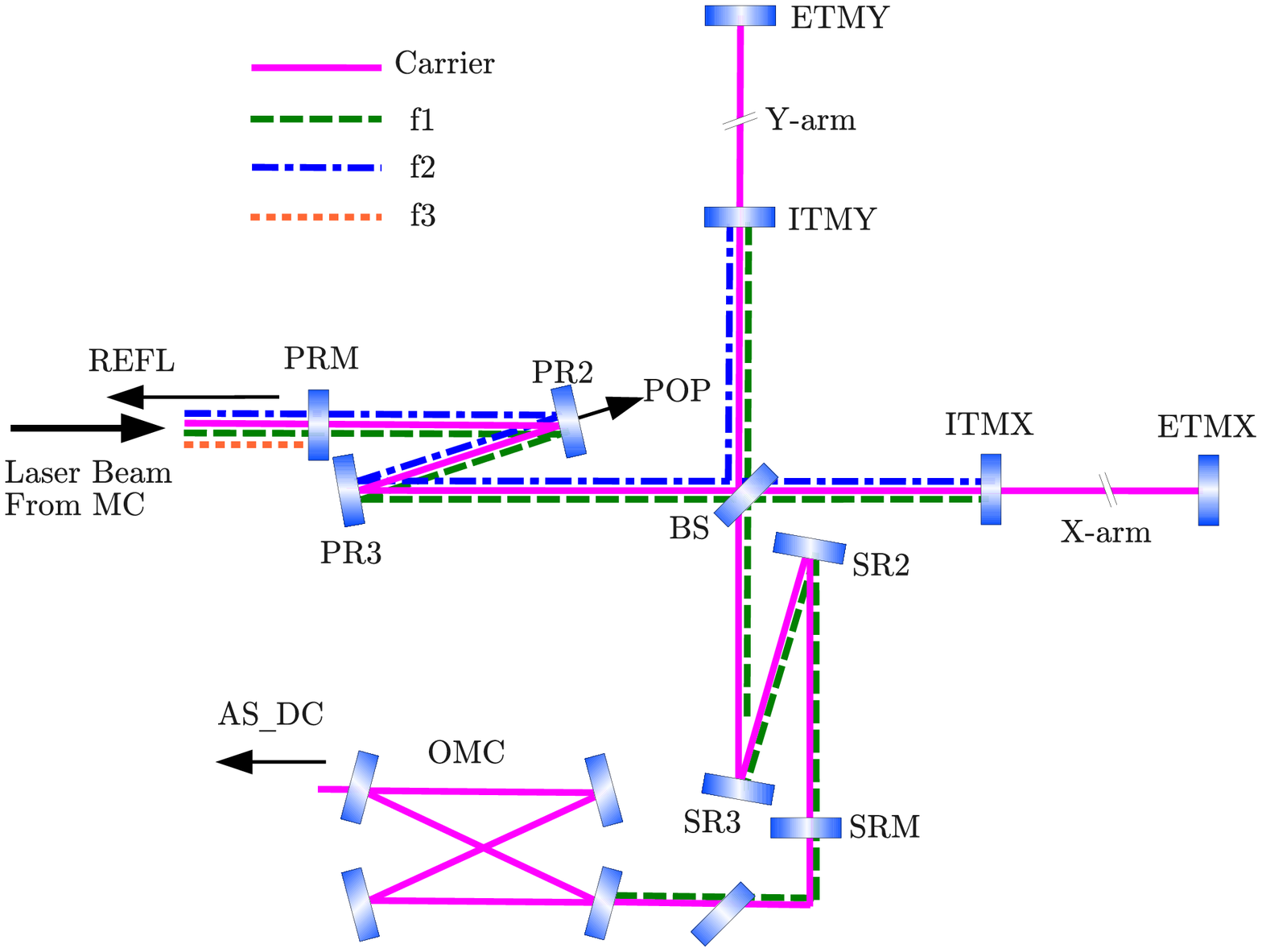} \caption{Interferometer
configuration and the sideband resonant conditions of LCGT. The laser
beam is injected through the mode cleaner (MC) from the left of the
figure. ETM: End Test Mass, ITM: Input Test Mass, BS: Beam Splitter,
PRM: Power Recycling Mirror, SRM: Signal Recycling Mirror, OMC: Output
Mode Cleaner. PR2, PR3, SR2 and SR3 are names of the folding
mirrors. REFL, POP and AS\_DC are names of the signal extraction ports.}
\label{Fig:InterferometerConfiguration}
\end{center}
\end{figure}

\begin{table}[tbp]
\begin{center}
\begin{tabular}{|c|c||c|c|}
\hline
Arm Cavity Finesse&1550&SRC Detuning for DRSE&3.5$^\circ$\\ \hline
Power Recycling Gain&10&Input Laser Power&51\,W\\ \hline
SRM Reflectivity&85\%&Main Mirror Mass&22.8\,kg\\ \hline
 \end{tabular}
\caption{Interferometer parameters of LCGT.}
\label{IFO Parameters}
\end{center}
\end{table}

LCGT will be constructed at the underground site of Kamioka mine with
the baseline length of 3\,km. The seismically quiet environment of the
Kamioka site is a great advantage for the stable and low noise operation
of sensitive devices like a gravitational wave detector.  Another major
feature of LCGT is the use of cryogenic sapphire mirrors, cooled down to
20K, to suppress the thermal noises.

The optical configuration of LCGT as well as the naming conventions of
the various parts of the interferometer are shown in
Figure\,\ref{Fig:InterferometerConfiguration}.  This configuration is
called power-recycled RSE. The basic parameters of the LCGT
interferometer are shown in Table~\ref{IFO Parameters}.  These
parameters are selected to maximize the scientific output of the
gravitational wave observation by optimizing the quantum noise spectrum
given practical constraints, such as the maximum cooling power for the
mirrors, and the spectra of classical noises\,\cite{Somiya_CQG_Amaldi9}.

LCGT is planned to be operated both in broadband RSE (BRSE) and detuned
RSE (DRSE) configurations. Therefore, the interferometer control scheme
has to be able to handle both the operation modes. Figure~\ref{Fig:LCGT
Noise} shows the projected noise curves of thee LCGT interferometer for
the two operation modes.

\begin{figure}[tbp]
\begin{center}
\begin{minipage}{7cm}
\begin{center}
  \includegraphics[width=7cm]{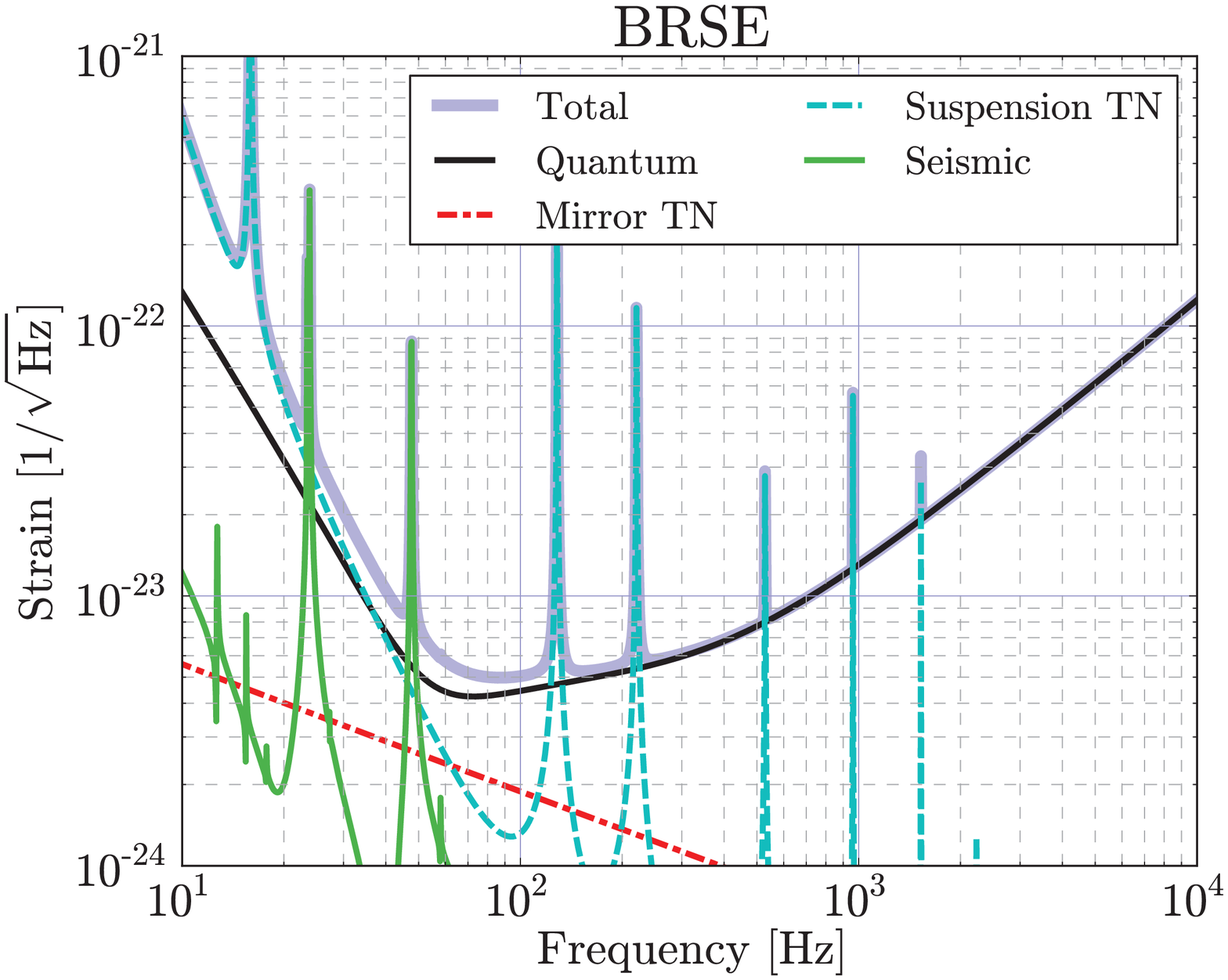}
\end{center}
\end{minipage}
\begin{minipage}{7cm}
\begin{center}
  \includegraphics[width=7cm]{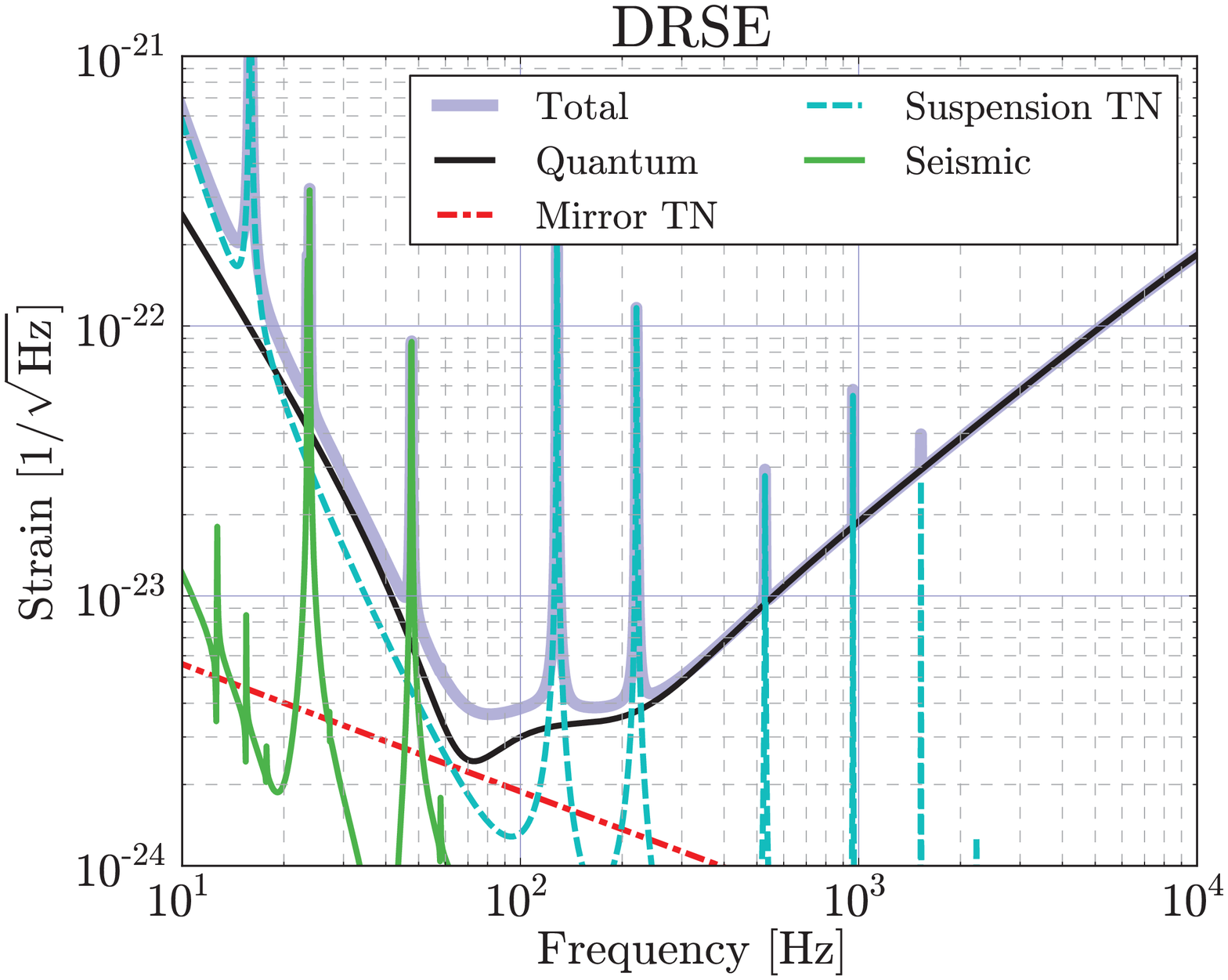}
\end{center}
\end{minipage}
\caption{Target sensitivities of the LCGT interferometer for the BRSE
 and DRSE operation modes.}
\label{Fig:LCGT Noise}
\end{center}
\end{figure}

 There is an output mode cleaner (OMC) at the downstream of the signal
 recycling cavity (SRC) to remove unwanted light components and only
 transmit the signal sidebands generated by the differential length
 change of the arm cavities. The signal sidebands contain the
 information of gravitational waves. They are converted into the power
 changes on a photo detector by the DC readout scheme, which is
 explained later.
 
The power recycling cavity (PRC) and SRC have
two mirrors each to fold the beam in a Z-shape. These folding parts serve
as telescopes to add extra Gouy phase changes in the cavities. The purpose
of the additional Gouy phase is to avoid the degeneracy of the
recycling cavities in terms of the higher order optical modes.

In order to keep the interferometer at the optimal operation point,
there are five length degrees of freedom to be controlled: i)
Differential change of the arm cavity lengths (DARM), ii) Common change
of the arm cavity lengths (CARM), iii) Differential change of the
Michelson arms formed by the BS and the two ITMs (MICH), iv) PRC length
(PRCL) and v) SRC length (SRCL). These five DOFs are represented as the
linear combinations of the mirror displacements, and called the
canonical DOFs in this paper.  DARM is the most important DOF, because it
contains information of gravitational waves. All other DOFs are collectively
called auxiliary DOFs. The purpose of a length sensing and control
system is to measure the variations of the canonical DOFs and apply
appropriate feedback to minimize the fluctuations.

%}}}

%{{{ \section{Signal Extraction Scheme}
\section{Signal Extraction Scheme}

%{{{ \subsection{RF Sidebands}

\subsection{RF Sidebands}

The main laser of LCGT is phase modulated at two radio frequencies
 (RFs), called f1 and f2, to generate RF sidebands. An additional
 amplitude modulation, called f3, is also applied during the lock
 acquisition. The resonant conditions of the RF sidebands are shown in
 Figure~\ref{Fig:InterferometerConfiguration}. The f1 sidebands are
 resonant both in the PRC and SRC, but not in the arm
 cavities. Therefore, they carry the information of the PRC and SRC,
 without affected by the arm cavities. Since the f1 sidebands partially
 transmit through the Michelson part, they are also sensitive to the
 MICH degree of freedom.  The f2 sidebands are resonant only in the
 PRC. Therefore, they are only sensitive to the change of PRCL. The
 amplitude modulated sidebands f3 are not resonant in any part of the
 interferometer, providing stable local oscillator fields for other RF
 sidebands. The f3 has to be amplitude modulation so that beating
 against other phase modulated sidebands yields zero-crossing error
 signals around the optimal operation point of the interferometer.

%}}}

%{{{ \subsection{Signal Extraction Ports}

\subsection{Signal Extraction Ports}

The light power coming out of the interferometer is detected at mainly
three ports: the reflection port (REFL), pick-off port in the PRC (POP)
and the anti-symmetric port (AS). REFL is the light coming back to the
laser, intercepted by a Faraday isolator. POP is taken at the
transmission of the PR2 to sample the light fields circulating in the
PRC. The AS port is the transmission of the OMC.

The light power fluctuations detected by photo detectors (PDs) at each
port are demodulated at various beat frequencies between the RF
sidebands and the carrier. Out of many possible combinations of the
signal ports, demodulation frequencies and demodulation phases (in-phase
or quadrature-phase), we selected the signals shown in the first columns
of Table~\ref{Sensing Matrix BRSE} and Table~\ref{Sensing Matrix DRSE}
to be used for the feedback control. The choice was made by repeatedly
computing the loop noise couplings, explained in section\,\ref{Section
Loop Noise}, with different combinations of the signal ports to find the
best one.

Because the f3 sidebands do not resonate in any part of the
interferometer, the demodulation of the REFL signal at the beat
frequencies of f3 and either f1 or f2 yields robust error signals for the
control of PRCL, SRCL and MICH (collectively called the central
part). Since those ``f3 signals'' do not depend on the carrier, they
have smaller couplings from CARM than the signals using the
carrier. Also the f3 signals are stable during the lock acquisition,
when the carrier rapidly changes its magnitude in the arm cavities.

Although the f3 signals have the advantages explained above, they can
have much worse shot noises than the signals using the carrier. First of
all, the modulation has to be strong enough to yield error signals with
small shot noise. However it will waste a lot of input laser power
unless some advanced modulation techniques are
used\,\cite{ohmae-phd}. At the REFL port, the amount of carrier light
strongly depends on the losses inside the interferometer, which is not
easily controllable. Since the carrier light is just a source of shot
noise for the f3 signals, this gives rise to an uncertainty in the
amount of shot noise. Therefore, we only plan to use the f3 signals
during the lock acquisition of the interferometer, not in the
observation mode.

Unlike the other DOFs, a scheme called DC readout\,\cite{ward_dc_2008}
is used to obtain the DARM error signal. A small DC offset is applied to
DARM to provide a local oscillator field for the DC readout. The
relative phase between the local oscillator and the signal sidebands
(homodyne angle) is set to 58$^\circ$ for BRSE and 45$^\circ$ for DRSE
to optimize the quantum noise level.

%}}}

%}}}

%{{{ \section{Macroscopic Lengths and Modulation Frequencies}

\section{Macroscopic Lengths and Modulation Frequencies}

%{{{ \subsection{Constraints}

\subsection{Constraints}
The macroscopic lengths of the PRC, SRC and the MICH asymmetry along
with the modulation frequencies of the RF sidebands have to be chosen to
satisfy the resonant conditions of
Figure~\ref{Fig:InterferometerConfiguration}. Also the length of the
mode cleaner (MC), which is used to clean the beam profile of the input
laser beam, has to be determined to transmit the RF sidebands through
it. In addition, various practical constraints are imposed on the choice
of those parameters.

The constraints on the available space is stringent in under ground
experiments like LCGT, where the space is at a premium.  In general we
want to minimize the lengths of the recycling cavities and the
MC. However, the recycling cavities have to accommodate the Z-shaped
folding part and the 20\,m long thermal radiation shields between BS and
ITMs. Therefore, the minimum possible recycling cavity length is about
65\,m.

Because a short MC has a larger free spectral range (FSR), a too short
MC will severely limit the choice of RF sideband frequencies, which have
to be integral multiples of the FSR. To strike a balance between the FSR
and the tunnel cost, the desirable MC length is in the order of 30\,m.

Another constraint is that the modulation frequencies have to be in the
range of 10\,MHz to 50\.MHz. If the frequency is too high, it is
difficult to find a good PD with fast enough response and a reasonably
large aperture.  If it is too low, the laser noises are not filtered out
enough by the pre-mode cleaner cavity.

%}}}

%{{{ \subsection{SRCL Linear Range}

\subsection{SRCL Linear Range}
There are many combinations of the macroscopic lengths and modulation
frequencies which satisfy the above constraints. In order to determine
the final parameter set to be used in the LCGT interferometer, we use
the linear range of the SRCL error signal as the figure of merit.

The operation mode is switched from BRSE to DRSE by adding an offset to
the SRCL error signal. The required detuning of the SRC is 3.5$^\circ$
in terms of one-way phase shift.  Therefore the SRCL error signal has to
have a large enough linear range to allow this offset.  This linear
range is roughly determined by the finesse of the coupled cavity formed
by the PRC and SRC for the f1 sideband. Since the reflectivities of the
PRM and the SRM are already determined by the optimization of the
quantum noise shape, we are left with the Michelson reflectivity for the
f1 sideband to change the finesse of the coupled cavity.

\begin{figure}[tbp]
\begin{center}
\includegraphics[height=5cm]{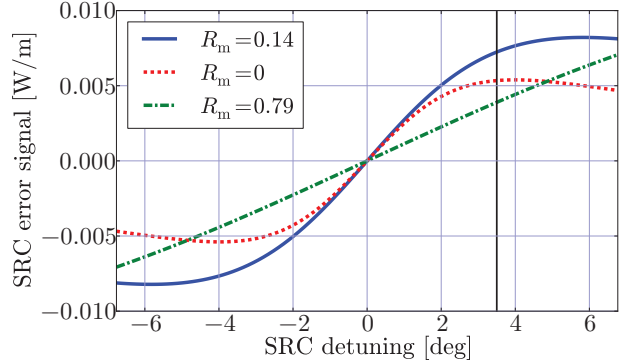}
\caption{SRCL error signals for three different values of
$R_\mr{m}$. The horizontal axis is the detuning of the SRC in terms of
the one-way phase shift. The vertical axis is the signal from the
POP port demodulated at the f1 frequency. The vertical line shows the
operation point of DRSE (3.5$^\circ$). } 
\label{Fig:SRCL Sweep}
\end{center}
\end{figure}

The Michelson reflectivity $R_\mr{m}$ depends on the f1 frequency
($f_1$) and the Michelson asymmetry ($l_\mr{m}$) as
$R_\mr{m}\propto\cos(2\pi f_1\cdot l_\mr{m} /c) $, where $c$ is the
speed of light. Figure\,\ref{Fig:SRCL Sweep} shows the shape of SRCL
error signals for three cases of $R_\mr{m}$.  When $R_\mr{m}$ is closer
to the PRM reflectivity (0.9), the effective reflectivity of the
power-recycled Michelson seen from the SRC becomes lower. Therefore, the
finesse of the SRC gets smaller, resulting in a wider linear range. In
the case of $R_\mr{m} = 0$, the DRSE operation point (shown by the
vertical line at 3.5$^\circ$) is almost at the turning point of the
error signal. Therefore it is not usable as an error signal. For
$R_\mr{m} = 0.79$, the error signal is linear throughout the plot range,
but the slope is smaller, meaning a poor shot noise. Therefore, after
an extensive parameter search, we decided to use a parameter set with
$R_\mr{m} = 0.14$, which has a larger slope at the center while the
signal is still not flat at 3.5$^\circ$.

The selected parameter set is shown in Table~\ref{Length Frequency
Parameters}. The f3 frequency is chosen to transmit the MC and not
resonant in any part of the interferometer including its higher order
harmonics.

\begin{table}[tbp]
 \begin{center}
{\small
  \begin{tabular}{|c|c||c|c|}
\hline
f1&16.875\,MHz&f2&45.0\,MHz \\ \hline
f3&56.3\,MHz&MC Length &26.6\,m\\ \hline
PRC Length&66.6\,m &SRC Length &66.6\,m \\ \hline
Michelson Asymmetry&3.33\,m&Michelson Reflectivity&0.14 \\ \hline
  \end{tabular}}
\caption{Macroscopic lengths and modulation frequencies of LCGT}
\label{Length Frequency Parameters}
 \end{center}
\end{table}
%}}}

%}}}

%{{{ \section{Loop noise couplings}

\section{Loop Noise Couplings}
\label{Section Loop Noise}

%{{{ \subsection{Interferometer Model}

\subsection{Interferometer Model}
In order to calculate the response and the quantum noises of the LCGT
interferometer, an interferometer model was constructed with an optical
simulation software called Optickle written by Matt
Evans\,\cite{evans_optickle_2007}.  The model uses the parameters
selected in the previous section. In order to simulate the imperfections of
real optics, 1\% asymmetries in the arm cavity finesse and the BS
reflectivity are introduced. These asymmetries increase off-diagonal
elements in the sensing matrices. They also create residual DC carrier
light at the AS port even without the DARM offset, allowing us to set
the homodyne angle to a desired value.

%}}}

%{{{ \subsection{Control Loop Modeling}

\subsection{Control Loop Modeling}
The sensing matrices computed by the interferometer model for the
canonical DOFs are shown in Table\,\ref{Sensing Matrix BRSE} and
Table\,\ref{Sensing Matrix DRSE}.  Because the off-diagonal elements in
the first row of the matrices are non-zero, there are finite couplings
from the auxiliary DOFs to the DARM error signal.

%{{{ Sensing Matrix Tables

\begin{table}[tbp]
 \begin{center}
{\small
  \begin{tabular}{|c@{\ \vrule width 0.8pt\ }c|c|c|c|c|}
\hline
&   {\bf DARM} &{\bf CARM} &{\bf MICH} &{\bf PRCL}&{\bf
   SRCL}\\\noalign{\hrule height 0.8pt}
   & & & & & \\[-10pt] 
{\bf AS\_DC}&1&$4.2\times10^{-5}$&$1.0\times10^{-3}$&$4.8\times10^{-6}$&$4.7\times10^{-6}$\\\hline
{\bf
   REFL\_f1I}&$5.4\times10^{-3}$&1&$4.3\times10^{-5}$&$6.5\times10^{-3}$&$4.3\times10^{-3}$\\\hline
{\bf REFL\_f1Q}&$5.0\times10^{-3}$&$1.3\times10^{-2}$&1&1.02&0.67\\\hline
{\bf
   POP\_f2I}&$2.3\times10^{-2}$&4.3&$1.0\times10^{-2}$&1&$2.5\times10^{-4}$\\\hline
{\bf POP\_f1I}&$8.7\times10^{-2}$&16.2&$3.1\times10^{-2}$&2.1&1\\\hline
  \end{tabular}}
\caption{Sensing matrix for the BRSE mode. Each row corresponds to a
  signal. AS\_DC is the DC readout at the AS port. Other signal names
  consist of a signal port (REFL or POP), demodulation frequency (f1 or
  f2) and demodulation phase (in-phase (I) or quadrature-phase
  (Q)). Each row is normalized by the diagonal element. The columns
  represent the canonical DOFs.  The n-th signal is fed back to the n-th
  DOF. The interferometer response was evaluated at 100\,Hz to create
  this matrix.}  \label{Sensing Matrix BRSE}
 \end{center}
\end{table}

\begin{table}[tbp]
 \begin{center}
{\small
  \begin{tabular}{|c@{\ \vrule width 0.8pt\ }c|c|c|c|c|}
\hline
&   {\bf DARM} &{\bf CARM} &{\bf MICH} &{\bf PRCL}&{\bf
   SRCL}\\\noalign{\hrule height 0.8pt}
   & & & & & \\[-10pt] 
{\bf AS\_DC}&1&$4.1\times10^{-5}$&$1.0\times10^{-3}$&$4.5\times10^{-6}$&$7.6\times10^{-6}$\\\hline
{\bf
   REFL\_f1I}&$1.2\times10^{-2}$&1&$1.3\times10^{-4}$&$1.2\times10^{-2}$&$1.4\times10^{-3}$\\\hline
{\bf REFL\_f1Q}&$2.8\times10^{-2}$&$9.9\times10^{-3}$&1&0.39&0.18\\\hline
{\bf
   POP\_f2I}&$2.7\times10^{-2}$&4.3&$1.0\times10^{-2}$&1&$8.5\times10^{-5}$\\\hline
{\bf POP\_f1I}&$1.7\times10^{-1}$&35&$3.1\times10^{-2}$&2.0&1\\\hline
  \end{tabular}
}
\caption{The sensing matrix for DRSE. The meanings of the rows and
  columns are the same as Table~\ref{Sensing Matrix BRSE}.}
\label{Sensing Matrix DRSE}
 \end{center}
\end{table}

%}}}

Sensing noises, most notably shot noise, are noise signals which do not
correspond to the real mirror motions. The feedback system tries to
cancel out these noises by unnecessarily moving the mirrors.  This
additional mirror motion is coupled to the DARM error signal through the
off-diagonal elements of the sensing matrix. The noise coupling of this
mechanism is called loop noise\,\cite{somiya_shot-noise-limited_2010}.

%{{{ Control Loop DIagram

\begin{figure}[tbp]
\begin{center}
\begin{minipage}{8cm}
\includegraphics[width=8cm]{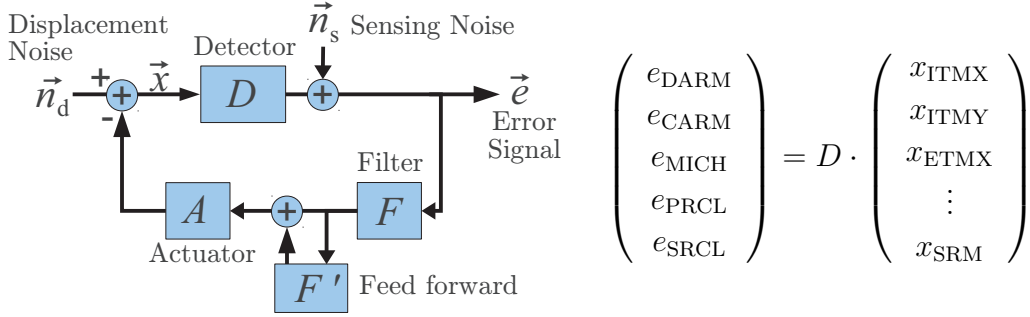}
\end{minipage}
\begin{minipage}{6cm}
$
 \left(
\begin{array}{c}
 e_\mr{DARM}\\
 e_\mr{CARM}\\
 e_\mr{MICH}\\
 e_\mr{PRCL}\\
 e_\mr{SRCL}
\end{array}
\right)
=D\cdot
\left(
\begin{array}{c}
 x_\mr{ITMX}\\
 x_\mr{ITMY}\\
 x_\mr{ETMX}\\
\vdots\\
 x_\mr{SRM}
\end{array}
\right)
$
\end{minipage}
\caption{Block diagram of the feedback loops. The mirror displacement
 vector $\vec{x}$ contains the displacement of each mirror. It is
 converted to the vector $\vec{e}$ of the error signals in the canonical
 DOFs by the detector matrix $D$. All the matrices in the figure are 
 frequency dependent.}  \label{Fig:Feedback Diagram}
\end{center}
\end{figure}

%}}}

The loop noise couplings can be modeled using the block diagram shown in
Figure\,~\ref{Fig:Feedback Diagram}. The detector matrix $D$ converts a
vector of mirror displacements $\vec{x}$ into a vector of error
signals $\vec{e}$ in the canonical DOFs. Then the sensing noise vector,
$\vec{n_\mr{s}}$, is added to the error signal vector. $D$ and
$\vec{n_\mr{s}}$ are calculated by the Optickle model. The error signals
are filtered by a feedback filter $F$ and fed back to the mirrors
through the actuator matrix $A$, which converts feedback signals in the
canonical DOFs to the motion of each mirror. The displacement noises of
the mirrors are represented by $\vec{n_\mr{d}}$.

The DARM error signal is the first element of the error signal vector
$\vec{e}$.  In the absence of gravitational waves, $\vec{e}$ is written
as,
\begin{equation}
\label{Loop Noise Formula}
 \vec{e} = (I+G)^{-1}\cdot \vec{n_\mr{s}} + (I+G)^{-1}\cdot
  D\cdot \vec{n_\mr{d}},
\end{equation}
\begin{equation}
\label{Define G}
 G\equiv D\cdot A\cdot (I+F')\cdot F,
\end{equation}
where $I$ is the identity matrix.  The off-diagonal elements of
$(I+G)^{-1}$ are responsible for the loop noise couplings.

% Ideally, the factor $(I+G)^{-1}$ in front of $\vec{n_\mr{s}}$ is
% diagonal. This means that only the sensing noise of the signal port for
% DARM (the DC readout port) contributes to the noise of the DARM error
% signal. However, in reality, this is not the case because of the
% off-diagonal elements in the sensing matrices.  Therefore, there is
% finite mixing of the shot noise from the auxiliary DOFs. Especially,
% MICH has an inherent coupling to DARM in the order of the inverse of the
% arm cavity finesse. The other auxiliary DOFs couple to DARM through
% their couplings to MICH. Although the couplings are small, since the
% shot noise of the auxiliary DOFs is much larger than DARM, these
% couplings can contaminate the DARM noise level.

%}}}

\begin{figure}[tbp]
\begin{center}
\begin{minipage}{7cm}
\begin{center}
\includegraphics[width=7cm]{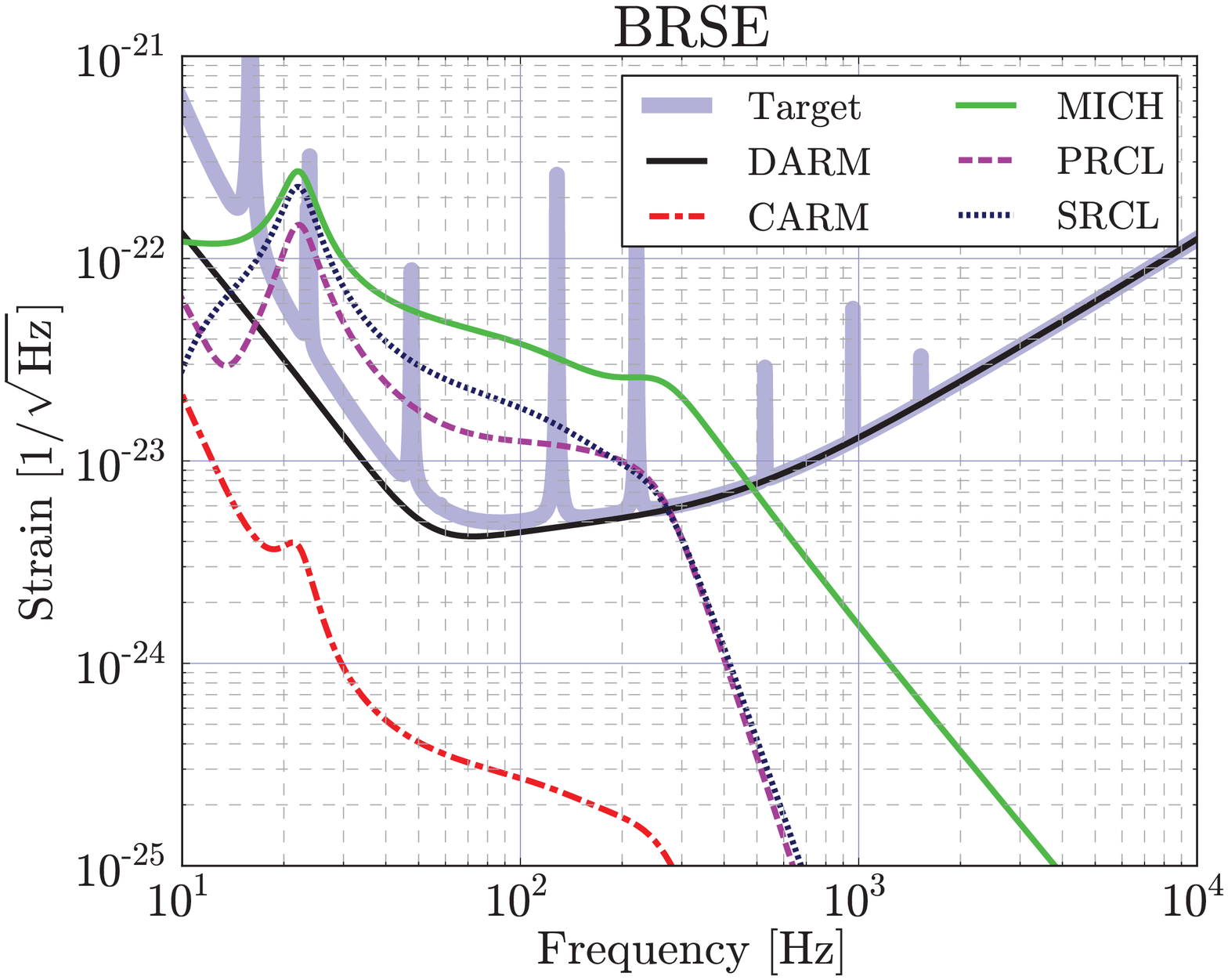}
\end{center}
\end{minipage}
\begin{minipage}{7cm}
\begin{center}
\includegraphics[width=7cm]{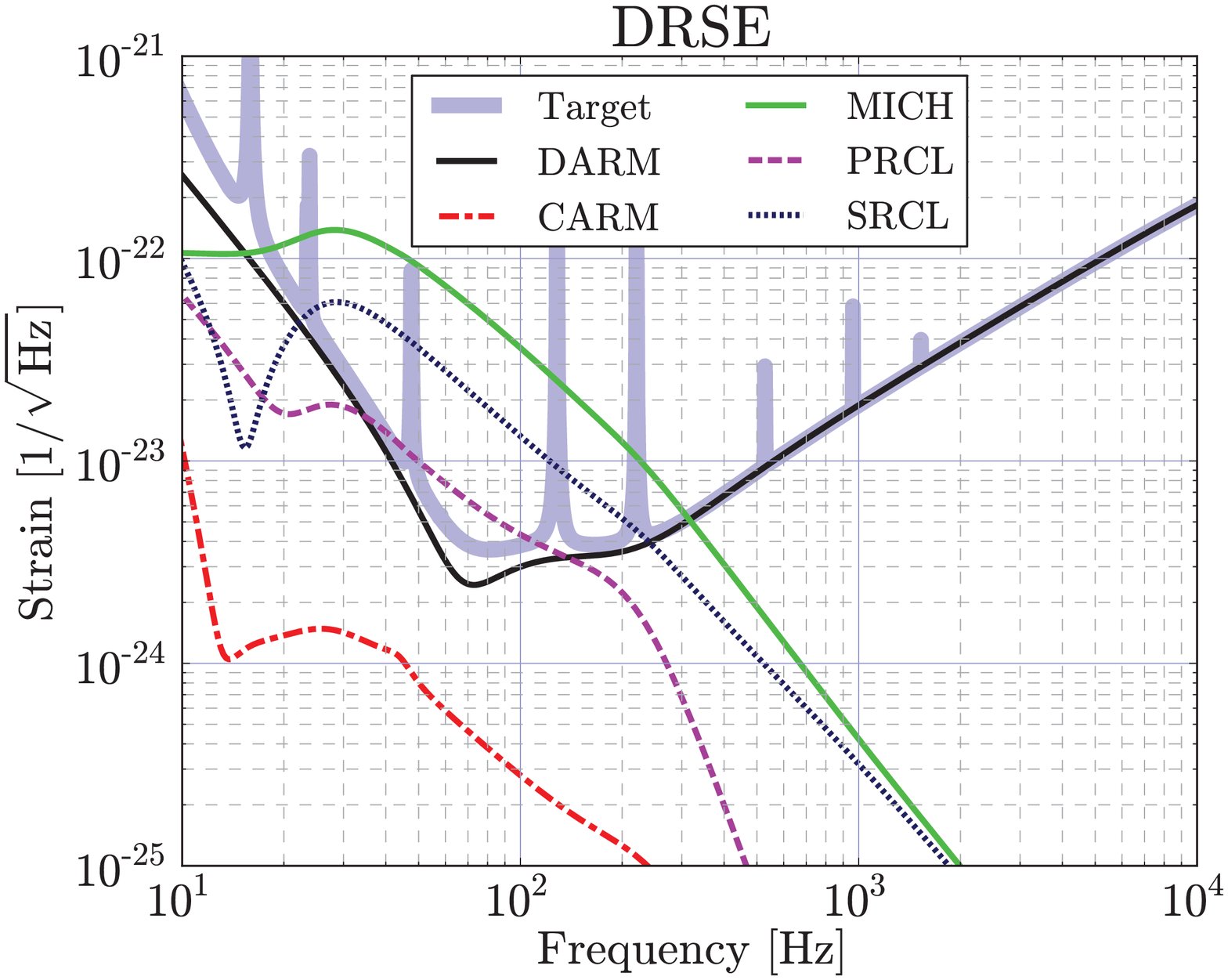}
\end{center}
\end{minipage}
\caption{Shot noise couplings to the DARM signal from the auxiliary
 DOFs} \label{Fig:Loop Noise}
\end{center}
\end{figure}

Figure~\ref{Fig:Loop Noise} shows the loop noise couplings from the
auxiliary DOFs to the DARM signal compared with the target sensitivity
of LCGT.  These noises are calculated by plugging in the shot noise of
each signal port to $\vec{n_\mr{s}}$ of (\ref{Loop Noise Formula}).  In
this calculation, we assumed simple $1/f^2$ shaped feedback filters with
$1/f$ response around the unity gain frequencies (UGFs) to ensure the
stability. The UGFs are 200\,Hz for DARM, 10\,kHz for CARM and 50\,Hz
for all the other DOFs. Clearly the shot noise couplings from the auxiliary DOFs
are larger than the target noise level.

%{{{ \subsection{Feed forward}

\subsection{Feed forward}
\begin{figure}[tbp]
\begin{center}
\begin{minipage}{7cm}
\begin{center}
\includegraphics[width=7cm]{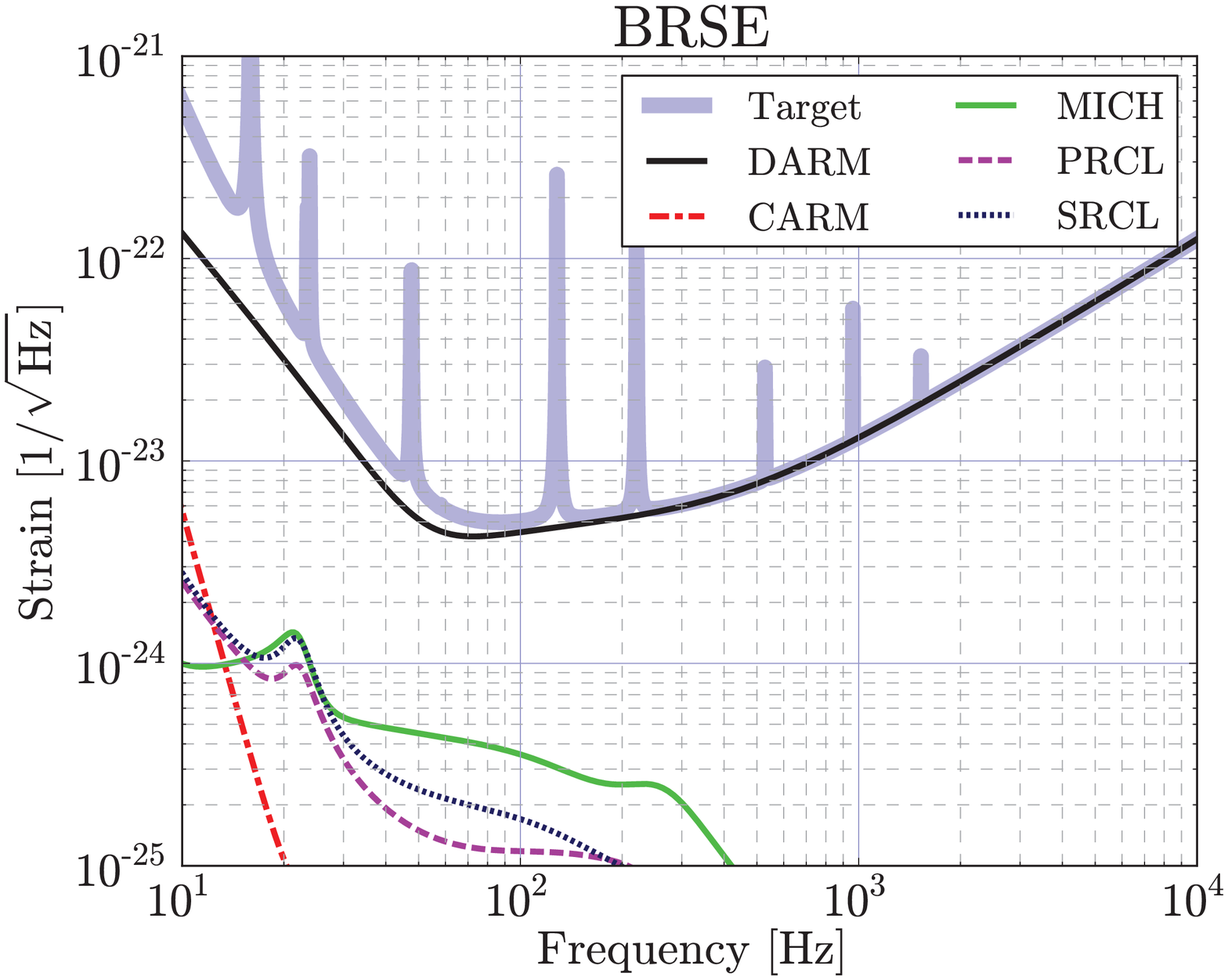}
\end{center}
\end{minipage}
\begin{minipage}{7cm}
\begin{center}
\includegraphics[width=7cm]{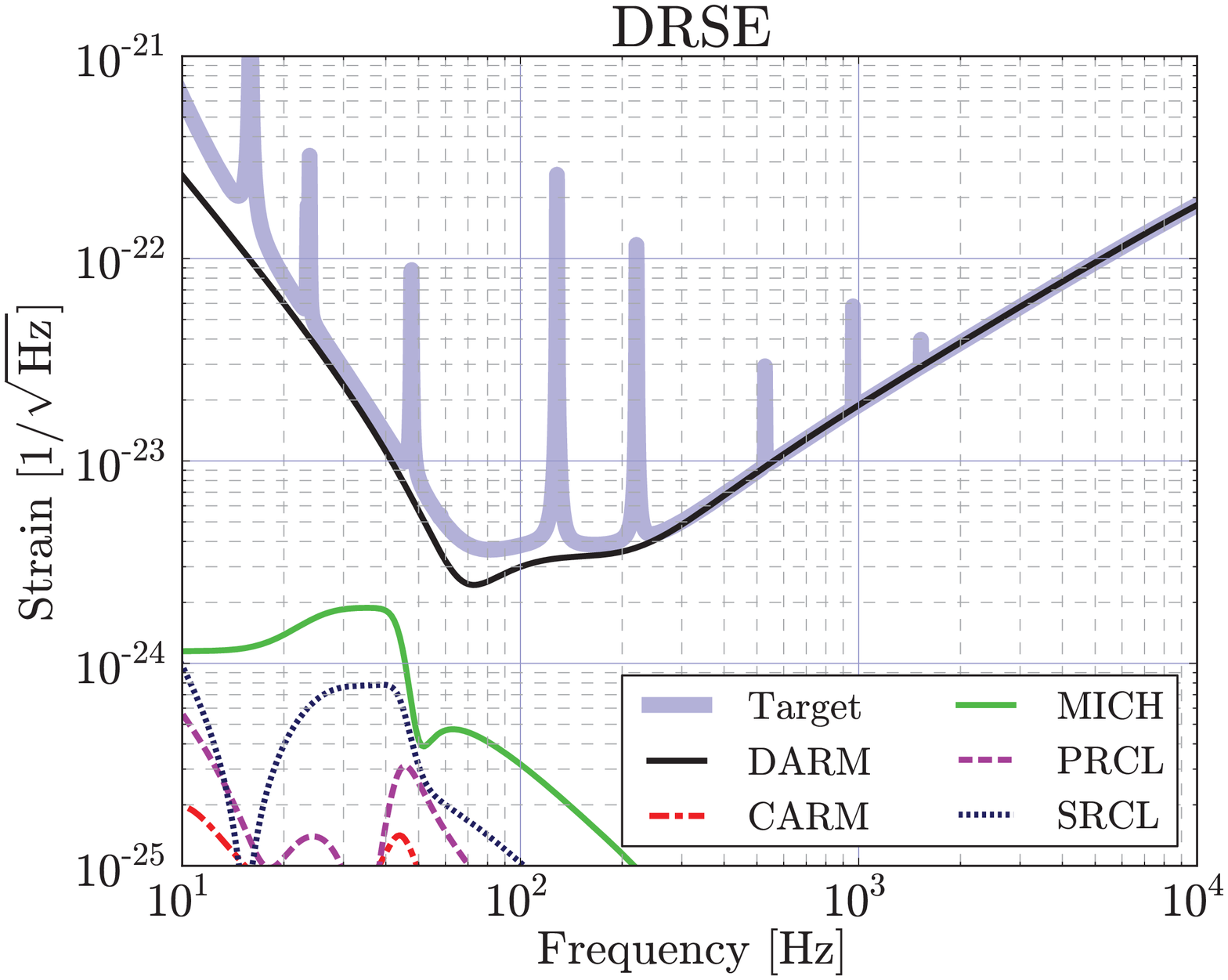}
\end{center}
\end{minipage}
\caption{Shot noise couplings with feed forward of gain 100}
\label{Fig:Loop Noise FF}
\end{center}
\end{figure}

The loop noise coupling problem can be mitigated by adding a feed
forward path $F'$ after the feedback filter $F$. The elements of $F'$
are determined by measuring the transfer functions from the actuation of
auxiliary DOFs to DARM error signal. The net effect of $F'$ is to reduce
the off-diagonal elements in the first row of $(I+G)^{-1}$.

The accuracy of the feed forward cancellation depends on the precision
of the transfer function measurements. The feed forward gain is defined
as the inverse of the error of the transfer function measurements. Thus
a 1\% error corresponds to a feed forward gain of 100. Since
the optical gains of the interferometer varies by the alignment
fluctuations, laser power variation and so on, the optimal feed forward
filters also change over time. Adaptive optimization of $F'$ is planed
to be used in LCGT.

Figure~\ref{Fig:Loop Noise FF} shows the quantum noise estimates when a
feed forward of gain 100 is applied. It is a reasonable assumption,
because the feed forward gain of more than 100 have achieved in the
first generation interferometers. In this case, all the shot noise
couplings from the auxiliary DOFs are well below the target noise level.

%}}}

%{{{ \subsection{Photo Detector Noise}

\subsection{Photo Detector Noise}
\begin{figure}[tbp]
\begin{center}
\begin{minipage}{7cm}
\begin{center}
\includegraphics[width=7cm]{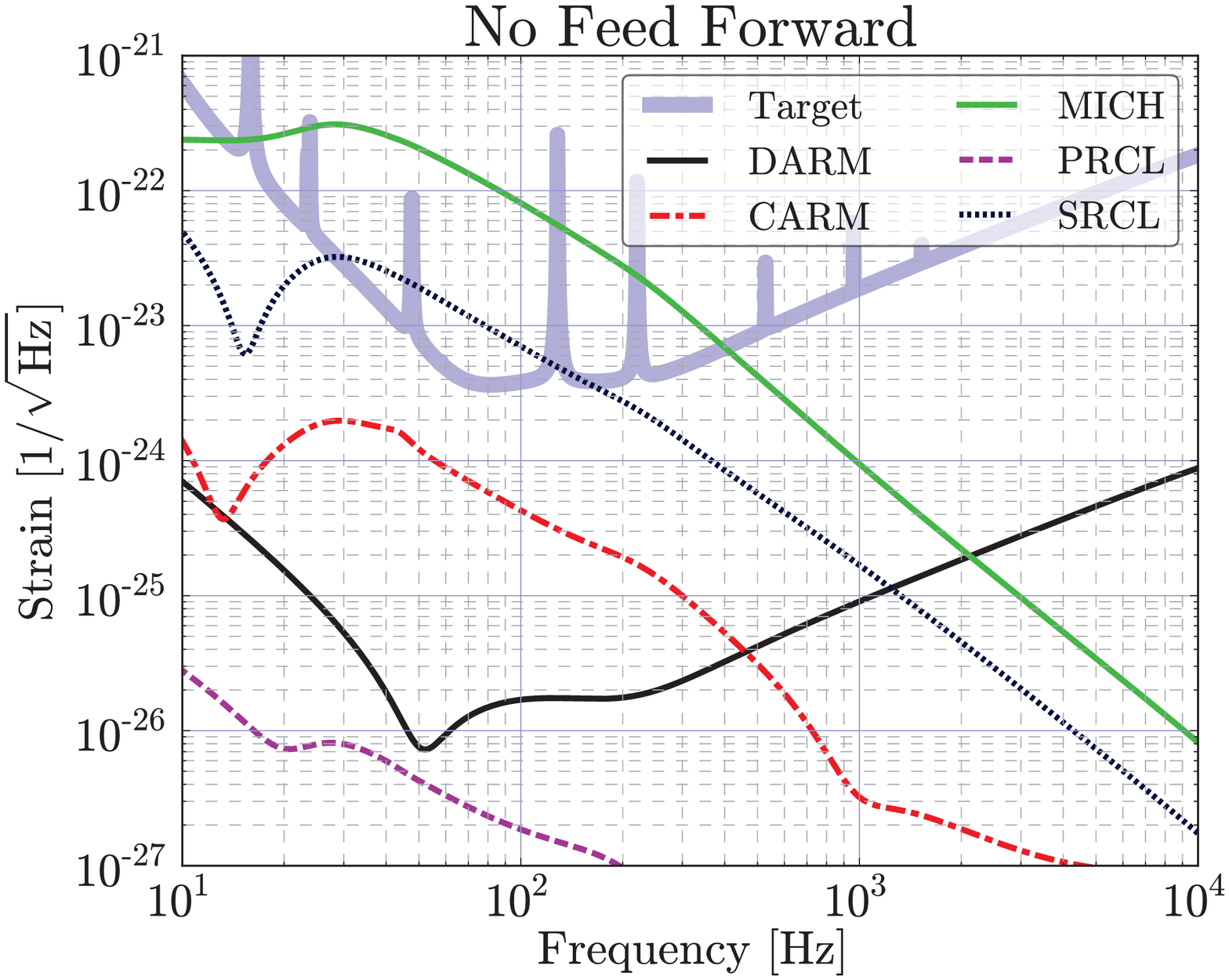}
\end{center}
\end{minipage}
\begin{minipage}{7cm}
\begin{center}
\includegraphics[width=7cm]{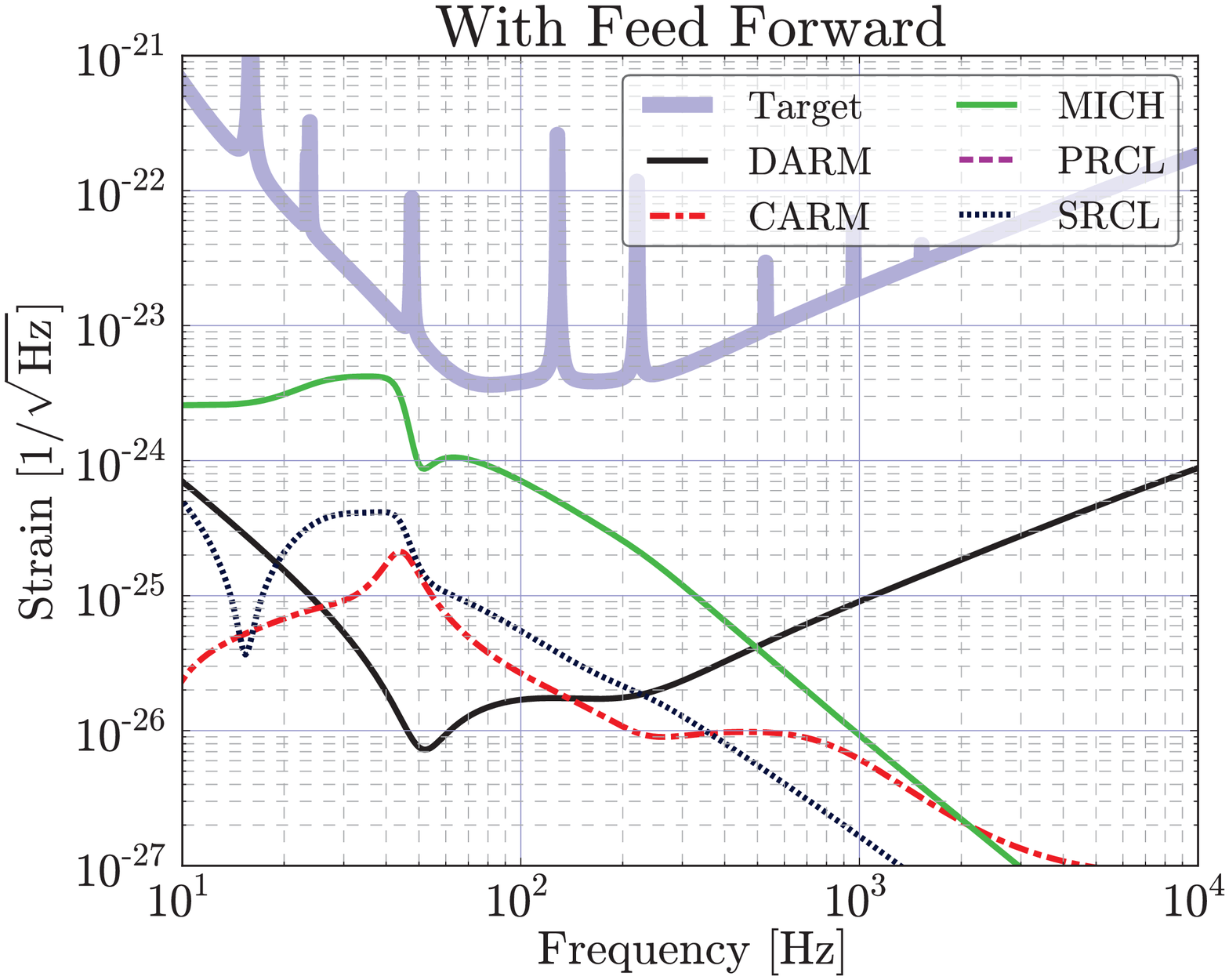}
\end{center}
\end{minipage}
\caption{PD noise couplings in the case of DRSE, with and without the
 feed forward.}
\label{Fig:PDNoise}
\end{center}
\end{figure}

Another important sensing noise to be considered is the intrinsic noises
of photo detectors.  A PD always receives some offset signal, either in RF or DC
depending on the type of PD. At some ports, these offset signals can be
very large. In this case, the dynamic ranges of the PD becomes an issue.

Typically, a low-noise fast operational amplifier (op-amp) used for the
current to voltage conversion of a PD has a dynamic range of about
200\,dB according to the catalog
specifications\,\cite{AD829_Data_Sheet}. However, because of the slew
rate limit, the actual dynamic range at RF is much smaller. Moreover, in
order to minimize the non-linearity of the detector response, we want to
use the op-amps at a much smaller signal level than the slew rate
limit. Therefore, for the following analysis, we assume the dynamic
range to be 160\,dB for RF PDs and 190\,dB for a DC PD.

Once the dynamic range $D$ is specified, the sensing noise, $n_\mr{pd}$,
of a PD, in terms of the equivalent signal light power on the PD, can be
expressed as $n_\mr{pd} = P_\mr{ofs}/D$, where $P_\mr{ofs}$ is the
offset signal power for the PD. Then we can simply replace
$n_\mr{s}$ in (\ref{Loop Noise Formula}) with $n_\mr{pd}$ to calculate
the loop noise couplings for the PD noise.

Figure~\ref{Fig:PDNoise} shows the calculated PD noise couplings in the
case of DRSE. The PD noises are large in the DRSE mode, especially for
MICH. It is because the SRC detuning changes the relative phase of the
f1 sidebands with the carrier so that they no longer form a pure phase
modulation.  The result is constant large RF signals on the 
PDs for the signals using the f1 sidebands. With the use of feed
forward, the PD noises can be made below the target noise level. For
BRSE (not shown in the figure), the PD noises are well below the target
noise even without the feed forward.

%}}}

%{{{ \subsection{Displacement noise requirement}

\subsection{Displacement noise requirements}
\begin{figure}[tbp]
\begin{center}
\begin{minipage}{7cm}
\begin{center}
\includegraphics[width=7cm]{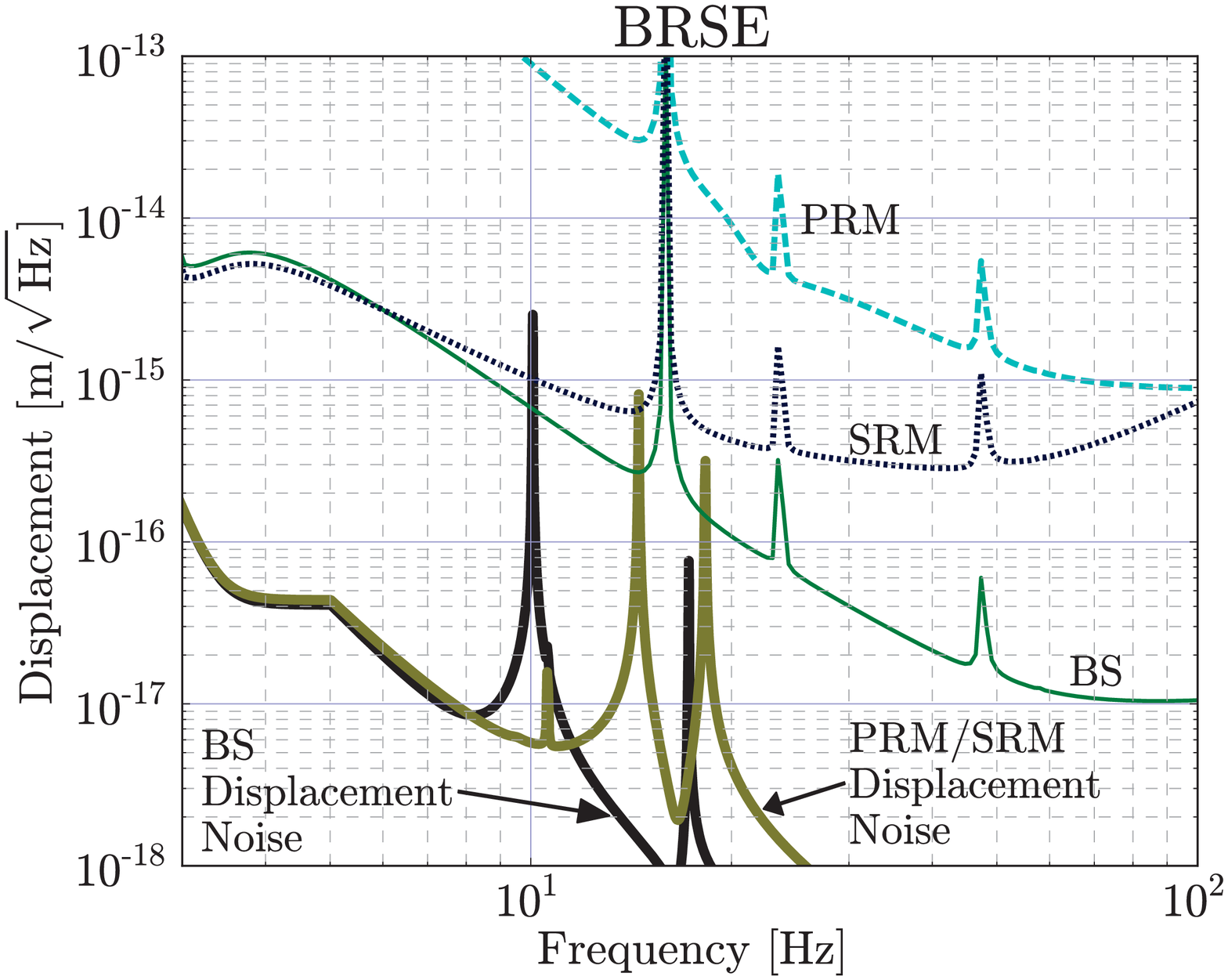}
\end{center}
\end{minipage}
\begin{minipage}{7cm}
\begin{center}
\includegraphics[width=7cm]{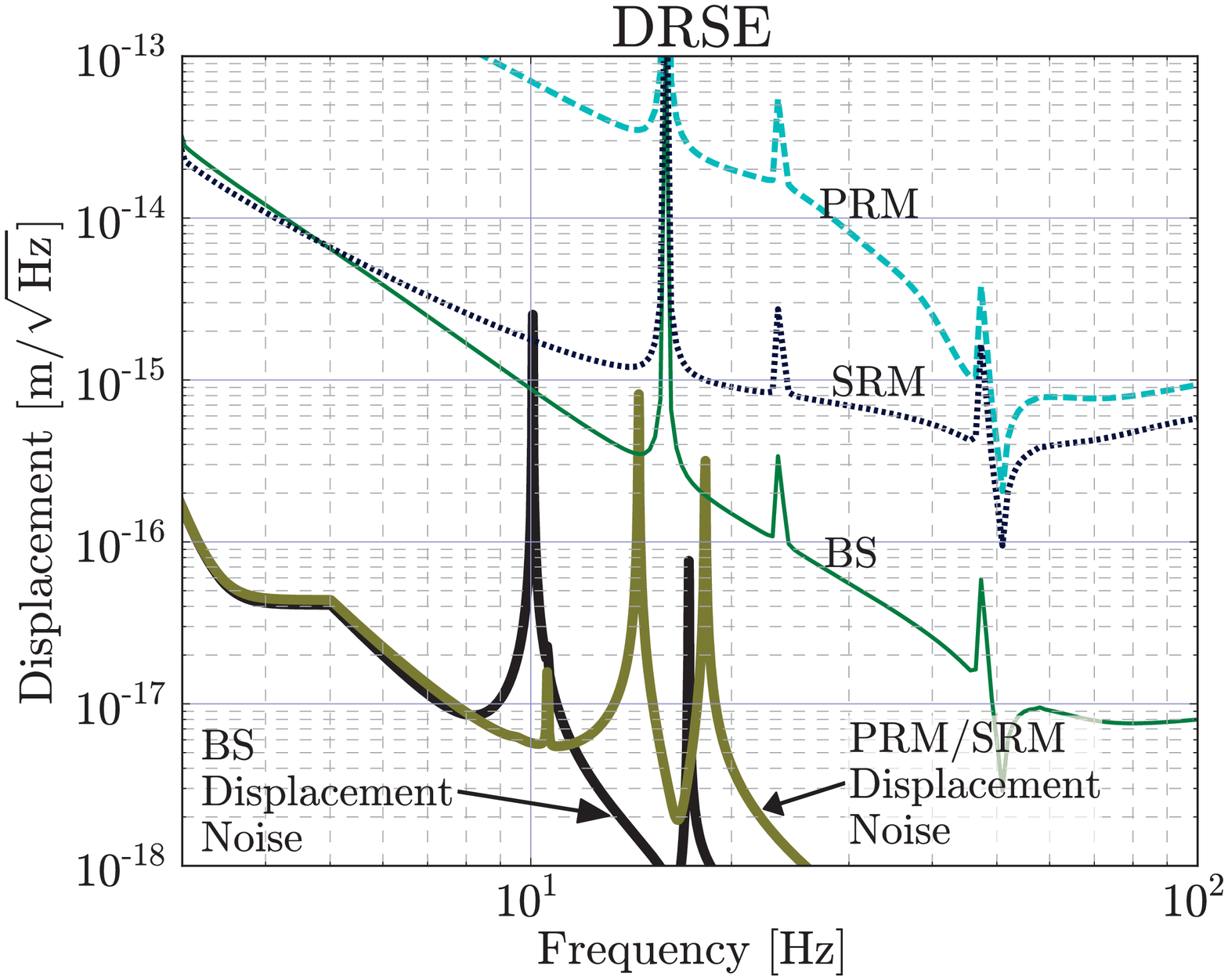}
\end{center}
\end{minipage}
\caption{Maximum permitted displacement noise for each mirror compared with
the estimated displacement noises.}
 \label{Fig:DispNoiseReq}
\end{center}
\end{figure}

The displacement noise contributions to the DARM signal can be calculated
from (\ref{Loop Noise Formula}) by plugging in the displacement noise of
each mirror to $\vec{n_\mr{d}}$. Conversely, the requirements to the
mirror displacement noise can be derived from (\ref{Loop Noise
Formula}) given a target noise level.  This is especially important when
feed forward is used, because feed forward is known to enhance the
displacement noise couplings\,\cite{somiya_shot-noise-limited_2010}.
To avoid this, the feed forward signals are cut off at
low-frequencies where displacement noises are dominant.

Figure~\ref{Fig:DispNoiseReq} shows the maximum permitted displacement
noise for each mirror requiring the contributions to the DARM signal be
smaller than the target noise level. In this calculation, feed forward
filters without the low-frequency cut off are assumed to show the worst
case scenario. The displacement noise estimates shown in the figure are
the sum of the seismic and thermal noises.  The seismic noise estimates
are based on the measured seismic data of very noisy day (during a heavy
storm) at the Kamioka site. Except for at several peaks, the expected
displacement noises are below the requirements in the observation band,
i.e. above 10\,Hz.

%}}}

%}}}

%{{{ \section{Lock acquisition}

\section{Lock acquisition}
\begin{figure}[tbp]
 \begin{center}
  \includegraphics[height=4cm]{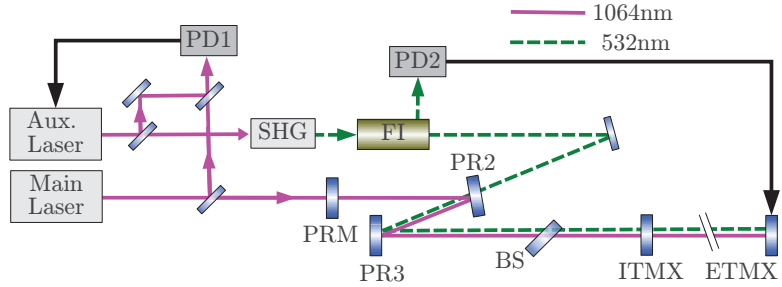}
\caption{A simplified schematic of the green laser pre-lock system. This
  diagram shows only the pre-lock system for the X-arm. There is a
  similar system for the Y-arm injected from the SR2.  FI: Faraday Isolator.}
  \label{Fig:GreenLock}
 \end{center}
\end{figure}

In order to ensure a high duty cycle during the observation and minimize
the turnaround time during the commissioning, a quick and robust lock
acquisition scheme is necessary. However, the relatively high finesse of
the arm cavities and the complex optical configuration of LCGT make it
difficult to achieve this. A high finesse cavity yields an useful error
signal only in a small region around the lock point. Even this signal can
be distorted by the slow transient response of the cavity if the mirrors
are moving fast. Moreover, the high optical power stored in
the arm cavities will kick the mirrors when the cavities are close to
the resonances, preventing the interferometer from gently settling down
to the locked state.
 
To assist the lock acquisition, we plan to use auxiliary lasers to
pre-lock the arm cavities. This idea was first developed at the 40\,m
laboratory in
Caltech\,\cite{Sam-Rana-Green-Lock}. Figure~\ref{Fig:GreenLock} shows a
conceptual diagram of the auxiliary laser lock system for the X-arm. An
auxiliary laser of 1064\,nm wavelength is phase locked with the main
laser. Then it is frequency doubled by a second harmonic generator (SHG)
to 532\,nm (green). The green laser beam is injected to the
interferometer from the back of the PR2. The coatings on the PR2 and BS
are designed to have a high transmissivity at 532\,nm while the PR3 has
a high reflectivity so that the beam is led only to the X-arm. The arm
cavity mirrors also have dichroic coatings to form a low finesse (about
10) cavity for the green light.  Because of the low finesse, it is easy
to lock the arm cavity with the green laser.  A similar system is also
prepared for the Y-arm.

Once the arm cavities are locked by the green lasers, the frequency
offset between the green and the main lasers is adjusted so that the arm
cavities are not resonant to the carrier nor any RF sidebands. This
ensures that the central part of the interferometer is not disturbed by
the arm cavities.

Error signals used to control the central part during the lock
acquisition should not depend on the carrier. Otherwise, the error
signals may be strongly disturbed when the arm cavities are finally
brought to the full resonances. Therefore, we will use the beat signals
between the f3 sidebands and the other RF sidebands to obtain the error
signals during the lock acquisition.  From Optickle simulations, it was
confirmed that these error signals are not noticeably affected by the
arm cavity offset from a carrier resonance.

After the central part is locked, the arm cavity offset is slowly
reduced by changing the offset frequency between the green and main
lasers. During this process, a large amount of optical power builds up
in the arm cavities, pushing the mirrors apart. This radiation pressure
is compensated by slightly leaning the suspension towers and letting the
gravity counter act the radiation pressure. Once the arm cavities reach
the full resonance, the control signals are handed off to the
observation mode signals: the DC readout for DARM and the RF signals
using the carrier for the auxiliary DOFs.

%}}}

%{{{ \section{Conclusion}

\section{Conclusion}
The length sensing and control scheme for the LCGT interferometer was
proposed. Optickle simulations show that the sensing noise couplings
caused by the control loops can be mitigated by the use of feed
forward. The displacement noise couplings to the gravitational wave
signal are confirmed to be less than the target noise level, even with
the enhancement by the feed forward.  For quick and robust lock
acquisition, auxiliary lasers will be used to pre-lock the arm
cavities. The control scheme explained here will be a key component in
the operation of the LCGT interferometer.

%}}}

%{{{ \section{Acknowledgements}
\section{Acknowledgements}
The authors would like to thank Matt Evans for the development of
Optickle. We would also like to thank Stefan Ballmer and Rana Adhikari
for helpful discussions. This work was supported by Grant-in-Aid for
Scientific Research (A) 22244049.
%}}}

%{{{ \section*{References}

\section*{References}
\bibliographystyle{iopart-num}\
\bibliography{master}

\providecommand{\newblock}{}
\begin{thebibliography}{1}
\expandafter\ifx\csname url\endcsname\relax
  \def\url#1{{\tt #1}}\fi
\expandafter\ifx\csname urlprefix\endcsname\relax\def\urlprefix{URL }\fi
\providecommand{\eprint}[2][]{\url{#2}}

\bibitem{mizuno_resonant_1993}
Mizuno J, Strain K~A, Nelson P~G {\em et~al.\/} 1993 {\em Physics Letters\/} A
  {\bf 175} 273--276

\bibitem{kuroda_status_2010}
Kuroda K and {The LCGT Collaboration} 2010 {\em Classical and Quantum
  Gravity\/} {\bf 27} 084004

\bibitem{Somiya_CQG_Amaldi9}
Somiya K an article to appear in this issue

\bibitem{ohmae-phd}
Ohmae N 2010 {\em Laser System for Second-Generation Gravitational-Wave
  Detectors\/} Ph.D. thesis University of Tokyo

\bibitem{ward_dc_2008}
Ward R~L, Adhikari R, Abbott B {\em et~al.\/} 2008 {\em Classical and Quantum
  Gravity\/} {\bf 25} 114030

\bibitem{evans_optickle_2007}
Evans M 2007 Optickle {LIGO} Document T070260

\bibitem{somiya_shot-noise-limited_2010}
Somiya K and Miyakawa O 2010 {\em Applied Optics\/} {\bf 49} 4335--4342

\bibitem{AD829_Data_Sheet}
For example, see the data sheet of AD829 available at http://www.analog.com/

\bibitem{Sam-Rana-Green-Lock}
Waldman S and Adhikari R 2007, personal communications

\end{thebibliography}

%}}}

\end{document}